\documentclass[12pt]{article}
\usepackage{epsfig}

\def\beq{\begin{equation}}
\def\eeq#1{\label{#1}\end{equation}}
\def\eeqn{\end{equation}}

\def\beqa{\begin{eqnarray}}
\def\eeqa#1{\label{#1}\end{eqnarray}}
\def\eeqan{\end{eqnarray}}

\let\bar=\overbar

\def\Dslash{\not{\hbox{\kern-4pt $D$}}}
\def\dslash{\not{\hbox{\kern-2pt $\del$}}}

\def\msb{{\bar{\ssstyle M \kern -1pt S}}}

\usepackage{fancyhdr,graphicx}
\def\Title#1{\begin{center} {\Large {\bf #1} } \end{center}}

\begin{document}

\Title{The influence of Strong Magnetic Field in Hyperonic Neutron Stars}

\bigskip\bigskip


\begin{raggedright}

{\it Luiz Laercio Lopes, Debora Peres Menezes \index{Vader, D.}\\
Universidade Federal de Santa Catarina\\
88040-900, Trindade  \\
Florian\'opolis, SC\\
Brazil\\
{\tt Email: luiz\_kiske@yahoo.com.br}}
\bigskip\bigskip
\end{raggedright}

\section{Introduction}

The physics of neutron stars leads historically towards  Landau's speculation. Even before the discovery of the neutron, he postulated
the possible existence of stars more compact than white dwarfs, containing matter of the order of  nuclear density~\cite{haensel}.
From a modern point of view neutron stars are compact objects maintained by the equilibrium between gravity and the degeneracy
pressure of the fermions together with a strong nuclear repulsion force due to the high density reached in their interior.
While the physics in the vicinity of nuclear saturation density is well know from  phenomenology, the physics of ultra-dense
 nuclear matter is still an open puzzle. In this work we study dense nuclear matter within a relativistic model, allowing
hyperons to be present through beta equilibrium. The presence of hyperons is justifiable since the constituents of neutron
stars are fermions. So, according to the Pauli principle, as the baryon density increases, so do the Fermi momentum and
 the Fermi energy~\cite{Glen}. On the other hand, this hyperonic matter softens the equation of state (EoS) and a recent
measurement of pulsar PSR J1614-2230~\cite{Demo} implies that the EoS has to be stiff enough to produce a 2.0 $M_{\odot}$
pulsar. We also consider  Duncan's magnetar ideas~\cite{Duncan} and study the influence of strong magnetic fields on the
EoS. We see that a strong magnetic field produces very massive neutron stars, in agreement with the astronomical observations.

\section{Formalism}

We describe the nuclear matter in beta equilibrium in the presence of a magnetic field with the non-linear Walecka model (NLWM)~\cite{Glen},
where the strong interaction is describe by the exchange of three massive mesons. The total Lagrangian density of the model is:

\begin{equation}
 \mathcal{L} = \sum_b \mathcal{L}_b + \sum_l \mathcal{L}_l + \mathcal{L}_m + \mathcal{L_B} , \label{1}
\end{equation}
where $b$  runs over the baryon octet ($N, H$), $l$ the two lightest leptons (e, $\mu$), $m$ the mesons ($\sigma$, $\omega$, $\rho$) and $\mathcal{B}$ the magnetic field itself.
Explicitly we have~\cite{bro,lopes}:

\begin{equation} 
\mathcal{L}_b = \bar{\Psi_b }[\gamma_u(i\partial^{\mu} - eA^{\mu} - g_{v,b}\omega^{\mu} - g_{\rho, b} I_{3b}  \rho^{\mu} ) - (M_b - g_{s,b} \sigma) ] \Psi_b , \label{2}
\end{equation}

\begin{eqnarray}
\mathcal{L}_m  = \frac{1}{2} \partial_\mu \sigma \partial^{\mu} \sigma - \frac{1}{2} m_s^2 \sigma^2 + \frac{1}{2} m_v^2\omega_\mu \omega^\mu - \frac{1}{4}\Omega_{\mu \nu}\Omega^{\mu \nu} +  \nonumber \\
+  \frac{1}{2} m_\rho^2 (\rho_\mu \rho^{ \mu}) - \frac{1}{4} {P}_{\mu \nu}  {P}^{\mu \nu}   -\frac{1}{3!}\kappa\sigma ^3 - \frac{1}{4!}\lambda\sigma ^4    \label{4} , \nonumber \\
\end{eqnarray}

\begin{equation}
\mathcal{L}_l =  \bar{\psi_l}[\gamma_u(i\partial^\mu - eA^{\mu})] - m_l]\psi_l , \label{3}
\end{equation}

\begin{equation}
\mathcal{L_B} = - \frac{1}{16 \pi}F_{\mu \nu} F^{\mu \nu} , \label{4e1}
\end{equation}
where $\Psi$ is the Dirac field. The g's are the coupling of the strong interaction with appropriate subscripts. The $\sigma$, $\omega$ and $\rho$ are
the mesonic fields, $A^\mu$ is the electromagnetic 4-vector field and $e$ is the electric charge. The second subscript of the g constants refers to the distinctive
 coupling of hyperons with the mesons. In this work, we assume two different values for the hyperon-meson coupling constants. One we call `Glendenning Conjecture' (GC),
 where all the hyperons are coupled with the same strength to a determined meson  $g_{s,H}=0.7g_{s,N}$; $g_{v,H}=0.783g_{v,N}$,
 and $g_{\rho,H} = 0.783g_{\rho,N}$~\cite{lopes,Glen2}, and another one, is the so called SU(6)~\cite{Weiss} parametrization, where:

\begin{equation}
 \frac{1}{3} g_{v,N} = \frac{1}{2} g_{v, \Lambda} = \frac{1}{2} g_{v \Sigma} = g_{v, \Xi} , \nonumber
\end{equation}
\begin{equation}
 \frac{1}{2} g_{\rho, \Sigma} = g_{\rho, N} = g_{\rho, \Xi} , \label{5} 
\end{equation}
\begin{equation}
 g_{\rho, \Lambda} = 0, \nonumber
\end{equation}
and the hyperon-sigma coupling is determined by the hyperon potential depths with: $U_\Lambda$ = -30 MeV, $U_\Sigma$ = +30 MeV and $U_\Xi$ = -18 MeV. We choose the 
parametrization of the bulk nuclear matter is given by the GM1 values~\cite{Glen,lopes,Glen2} as presented in Table 1\footnote{For a study with GM3
 parametrization see~\cite{lopes2}.}.

\begin{table*}[ht]
\centering

\caption{Values of GM1 parametrization.}
\label{tab:1}       
\begin{tabular}{llllll}\hline\noalign{\smallskip}
Set & $(g_s/m_s)^2$ & $(g_v/m_v)^2$ & $(g_\rho/m_\rho)^2$ & $\kappa/M_N$ & $\lambda$  \\
\noalign{\smallskip}\hline\noalign{\smallskip}
GM1 & 11.785 $fm^2$& 7.148 $fm^2$ & 4.410 $fm^2$ & 0.005894 & -0.006426 \\
\noalign{\smallskip}\hline
\end{tabular}
\vspace*{1cm}  
\end{table*}

To obtain the EoS associated with the NLWM, we use a mean field approximation, where the meson fields are replaced by their expectation values\footnote{For
a detailed calculation see~\cite{Glen,lopes,Peng}.}.
The energy density of the system described by the above Lagrangian density (\ref{1}) reads:

\begin{equation}
\epsilon  =  \sum_{ub}\epsilon_{ub} + \sum_{cb}\epsilon_{cb} + \sum_l \epsilon_l + \sum_m \epsilon_m + \frac{B^2}{8\pi} , \label{22}
\end{equation}
where $ub$ stands for ``uncharged baryons'', $cb$ for ``charged baryons'', $l$ for the ``leptons'', $m$ for the ``mesons'' and the last term is the contribution of
the magnetic field itself. Explicitly:
\begin{equation}
\epsilon_{ub} = \frac{1}{\pi^2}\int_{0}^{k_f} \sqrt{M^{*2}_b + k^2}k^2 dk, \label{23}
\end{equation}
\begin{equation}
\epsilon_{cb} = \frac{|e|B}{2\pi^2} \sum_{\nu}^{\nu_{max}}\eta(\nu) \int_{0}^{k_f} \sqrt{ M^{*2}_b +  k_z^2 +2\nu |e|B}dk_z \label{24} ,
\end{equation}
\begin{equation}
\epsilon_{l} = \frac{|e|B}{2\pi^2} \sum_{\nu}^{\nu_{max}}\eta(\nu) \int_{0}^{k_f} \sqrt{ m^{2}_l +  k_z^2 +2\nu |e|B}dk_z \label{25} ,
\end{equation}
\begin{equation}
\epsilon_{m} = \frac{1}{2}m_s^2\sigma_0^2 + \frac{1}{2}m_v^2\omega_0^2 + \frac{1}{2}m_\rho^2\rho_0^2 + \frac{1}{3!}\kappa \sigma_0^3 + \frac{1}{4!}\lambda \sigma_0^4 \label{26} .
\end{equation}
The discrete parameter $\nu$ is called Landau level ($LL$). The first $LL$ ($\nu$ =0) is non-degenerate and all the others are two-fold degenerate. The maximum 
allowed Landau level is the largest value of $\nu$ for which the square of  Fermi momentum of the particle is still positive.

The number densities are given by:

\begin{equation}
n^{ub} =   \int_{0}^{k_f} \frac{8\pi k^2}{(2\pi)^3} = \frac{k_f^3}{3\pi^2}, \label{15}
\end{equation} 
\begin{equation}
n^{cb} =  \frac{|e|B}{(2\pi)^2}\sum_{\nu}^{\nu_{max}}\eta{(\nu)} \int_{-k_f}^{k_f}dk_z = \frac{|e|B}{2\pi^2}\sum_{\nu}^{\nu_{max}}\eta (\nu)k_f \label{16} .
\end{equation}

To find the pressure, we use the second law of thermodynamics that gives an isotropic pressure:

\begin{equation}
p = \sum_i \mu_i n^{i} -  \epsilon + \frac{B^2}{8\pi} \label {27} ,
\end{equation}
where the sum runs over all fermions. Note that the contribution from the electromagnetic field should be taken into account in the calculation of
 the energy density and the pressure.

The magnetic field at the surface of the magnetars are of order of $10^{15}G$, but can reach more than $10^{18}G$ in their cores.
 To reproduce this behaviour we use a density-dependent magnetic field as given in~\cite{lopes}, and use the TOV equation~\cite{TOV}
to obtain the microscopic properties of a relativistic star.

\section{Results,  discussion and conclusions}

We choose two values for the magnetic field: $1.0 \times 10^{17}G$  and $3.1 \times 10^{18}G$ to produce a weak and a strong influence.
 The strongest value used here is the maximum allowed value that keeps the EoS in an isotropic regime~\cite{XH}. Now we define the  particle fraction
 $Y_i = n_i /n$ and plot them with the GC (top) and the SU(6) (bottom) parametrizations subject to a weak (left) and strong (right) magnetic field  in Fig. 1.
 
The main difference between the SU(6) and GC coupling is the absence of the triplet $\Sigma$ in the first case. This is a very drastic difference,
 since the $\Sigma^-$ is the first hyperon to appear if the GC parametrization is used. Moreover, all the triplet is present in a non-negligible fraction with
 the GC coupling.
On the other hand the absence of the $\Sigma$-triplet promotes a more significant fraction of the doublet $\Xi$. Indeed the $\Xi^-$ now appears much
 earlier than in the EoS with the GC coupling.
 Also the $\Xi^0$ that is absent with GC, has now a prominent presence in SU(6). This is due to the repulsive potential depth felt by the $\Sigma$,
 and the very weak coupling to the $\Xi$ with the vector meson $\omega$ in SU(6). We see that a strong magnetic field causes the appearance of many kinks.
 This is due to the Landau quantization. For a weak magnetic field a lot of Landau levels are occupied, and the result
approaches the Fermi distribution, as can be seen for the zero magnetic field in~\cite{Glen,Weiss}. But for a strong magnetic field, just a few of
 them are filled. For strong magnetic fields, the appearance of charged particles is favoured at low densities due to their dependence on the magnetic field, as
shown in eq. (\ref{16}).
The effect of the magnetic field and the two different parametrizations can be seen clearer when we plot the total strangeness fraction defined as:

\begin{equation}
 f = \frac{1}{3} \frac{\sum_j |s_j| n_j}{n} ,
\end{equation}
where $s_j$ is the strangeness,  $n_j$ is the number density of the baryon j and $n=\sum n_j$ is the total
number density. We plot the results in Fig.2.
We see that although the magnetic field causes many kinks in the individual $Y_i$, the total strangeness fraction washes out this effect. The best outcome
arises when we compare both parametrizations. With the GC parametrization $f$ starts earlier than with the SU(6) one. However, the SU(6) parametrization,
 $f$ grows quicker than with the GC one, assuming larger values at higher densities. This happens because with the GC
parametrization, the $\Sigma^-$ appears very soon, and strangeness appear at relatively low densities. Within the SU(6),
the hyperons couple very weakly to the vector meson, forcing the total strangeness fraction to be larger at higher densities.

 \begin{figure}[ht]
 \centering{Fraction of particles $Y_i$ as a function of number density $n$.}
\begin{tabular}{cc}
\includegraphics[width=6.4cm,height=7.2cm,angle=270]{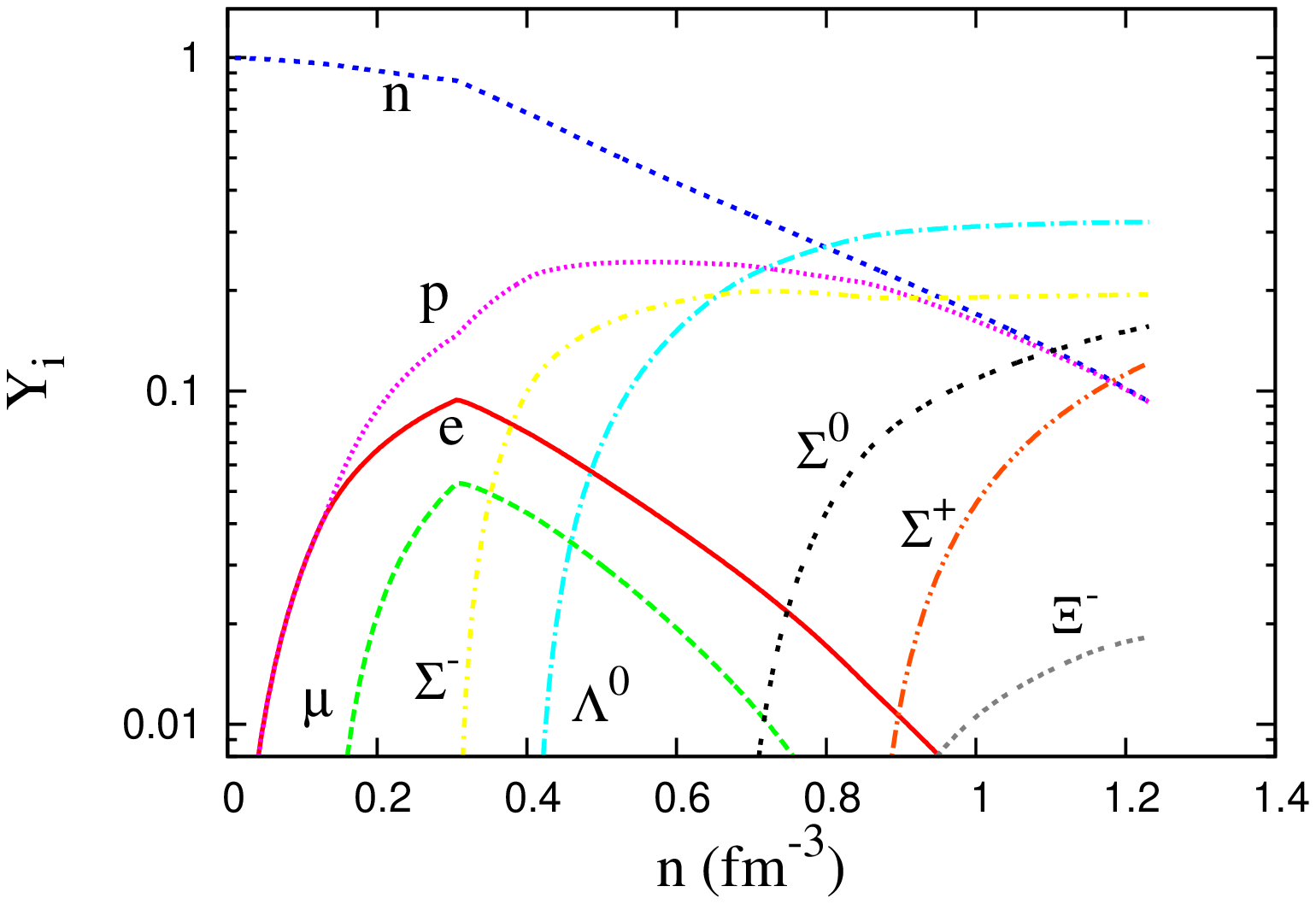} &
\includegraphics[width=6.4cm,height=7.2cm,angle=270]{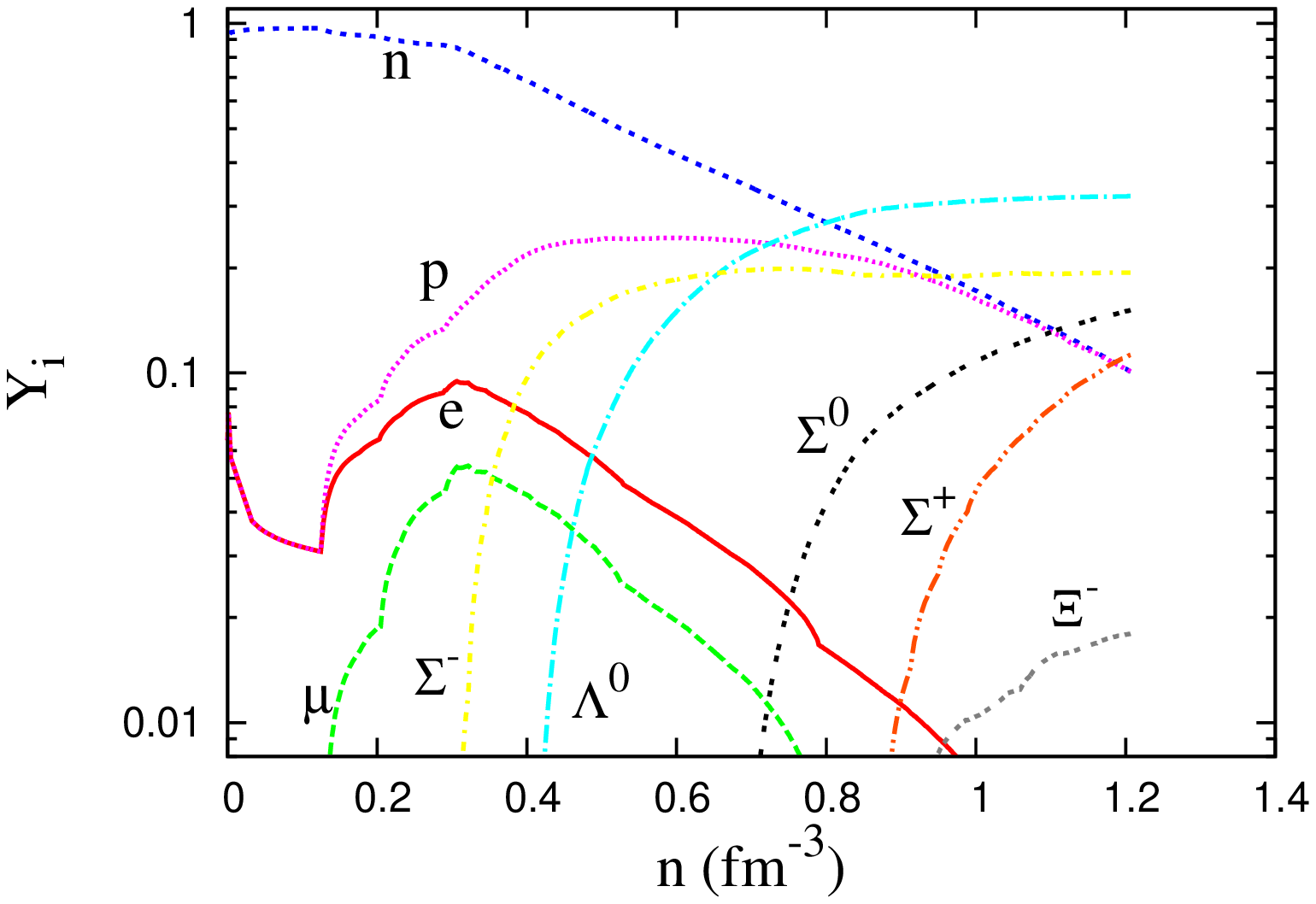} \\
\includegraphics[width=6.4cm,height=7.2cm,angle=270]{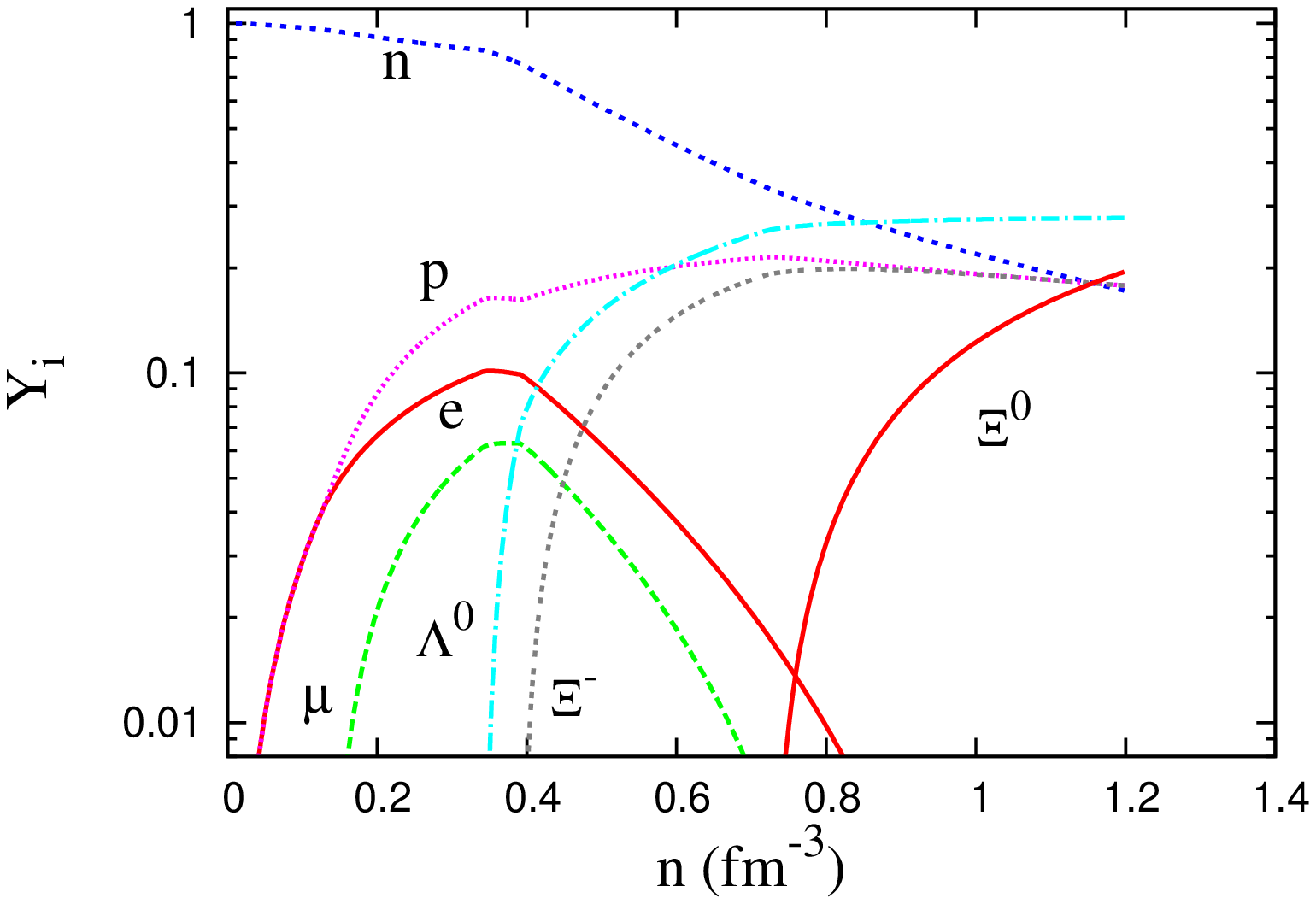} &
\includegraphics[width=6.4cm,height=7.2cm,angle=270]{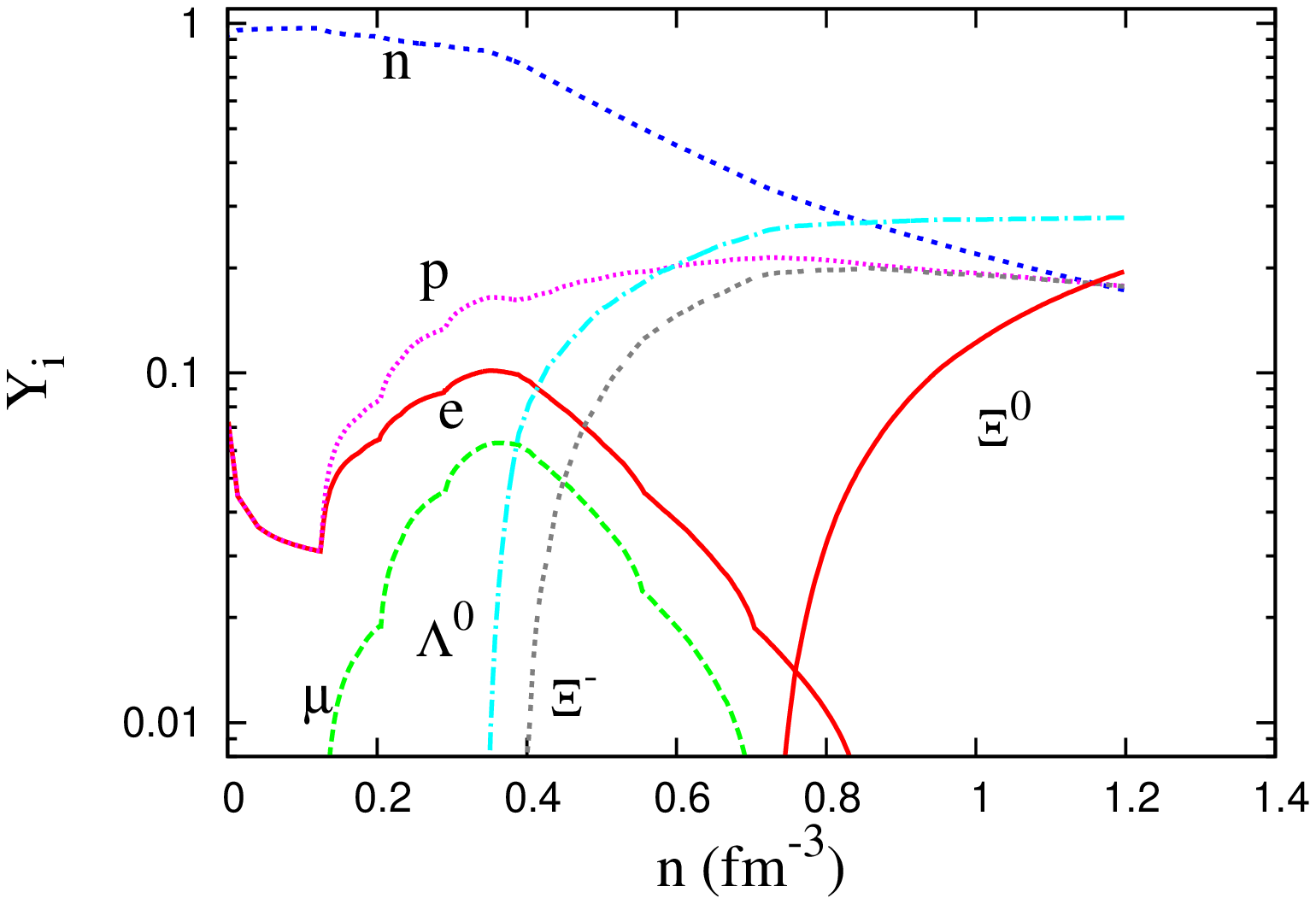} \\
\end{tabular}
\caption{ (Top) GC coupling  for a magnetic field of $1.0\times 10^{17}G$ (left) and 3.1$\times 10^{18}G$. (Bottom) The same for the SU(6).}
\end{figure}

\noindent 

Now we calculate the EoS for these four cases discussed above and obtain the corresponding TOV mass-radii solutions and plot the results in Fig 3.
 On the  left we have the  EoS, the ``c'' line is the causal limit,  and on the right  the TOV solution, where the hatched area is the uncertainty in the mass
 of the Demorest pulsar~\cite{Demo}. 

\begin{figure}[ht]
\centering
\includegraphics[angle=270,
width=0.5\textwidth]{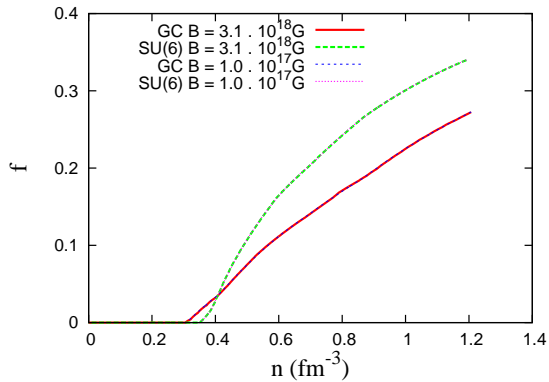}
\caption{ Total strangeness fraction $f$.}
\end{figure}
 \begin{figure}[ht]
\begin{tabular}{cc}
\includegraphics[width=6.cm,height=7.cm,angle=270]{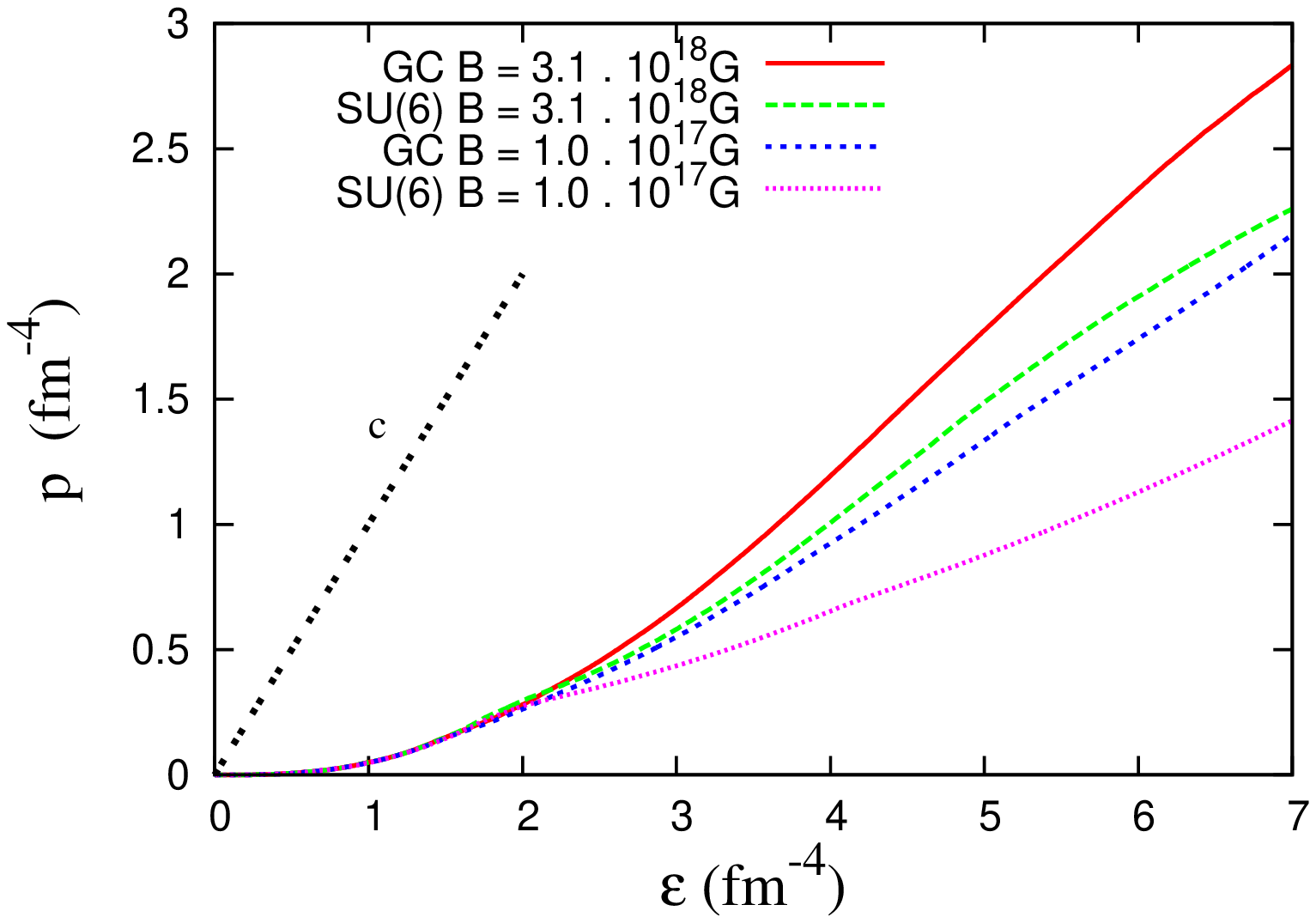} &
\includegraphics[width=6.cm,height=7.cm,angle=270]{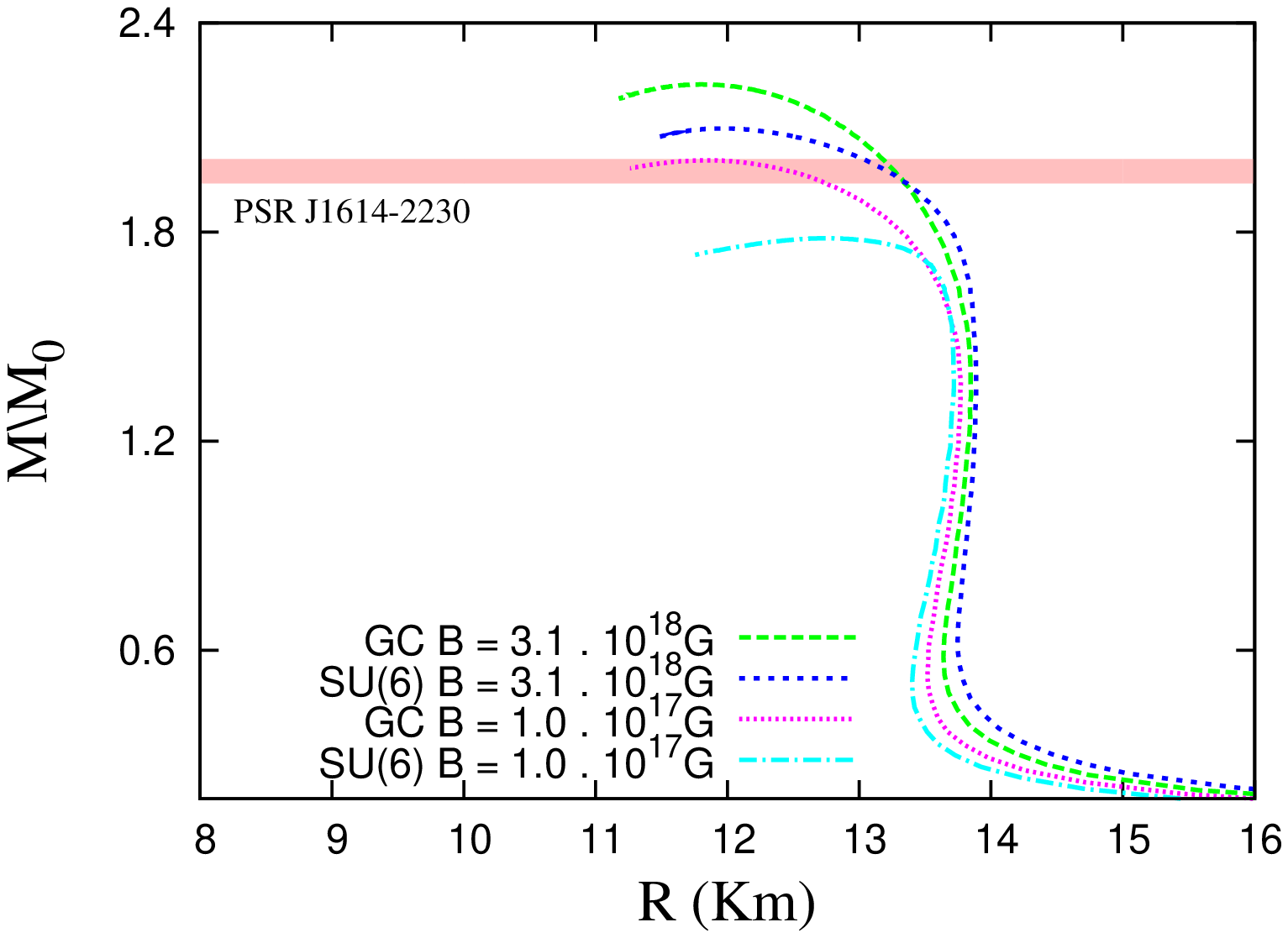} \\
\end{tabular}
\caption{ (Left) EoS for the four cases discussed in the text.  (Right)  The same for the TOV solutions.}
\end{figure}

We see that the harder the EoS, the more massive is the neutron star produced.We also see that although the SU(6) parametrization
cannot reproduce the Demorest pulsar for a weak magnetic field, if it is a magnetar with strong magnetic field, that parametrization is able to predict it.
We also may wonder what is the lowest possible mass in such a way that the strangeness fraction has a considerable value, let's says 5\%.
These results are shown in Table 2, where we see that no neutron star with mass below 1.65 $M_\odot$ has a significant strangeness fraction in its composition.

\begin{table}[ht]
\centering
\begin{tabular}{rrrrrr}
\hline
\hline
 {\bf Family}  & {\bf B ($\times 10^{18}G$)} & {\bf $M/M_{\odot}$} &  {\bf R $(Km)$} &{\bf n $(fm^{-3})$} & {\bf f}  \\
\hline
\hline
SU(6) & 0.1 & 1.78 & 12.75 & 0.787  & 0.24 \\
GC & 0.1 & 2.01 & 11.86 & 0.952 & 0.21  \\
SU(6) & 3.1 & 2.09 & 11.95 & 0.590 & 0.16\\
 
GC & 3.1 & 2.22 & 11.80 &0.842 & 0.18\\
\hline
SU(6) & 0.1 & 1.65 & 13.62 & 0.424  & 0.05\\
GC & 0.1 & 1.70 & 13.52 & 0.442 & 0.05\\
SU(6) & 3.1 & 1.74 & 13.74 & 0.424 & 0.05\\
 
GC & 3.1 & 1.81 & 13.56 &0.443 &  0.05\\
\hline
\hline
\end{tabular}
\caption{(Top) Maximum mass for the four cases discussed in the text. (Bottom) Lowest mass in such a way that $f=0.05$ }
\label{tab:a}
\end{table}

We conclude our work noting that the SU(6) not only predicts less massive pulsars, but also less dense ones. The opposite is found when we use the GM3 parametrization
~\cite{lopes2},  for which the least massive pulsar is the  denser one. We also reinforce the fact that the magnetic field  has to be taken into account in
the description of massive magnetars, since they can increase the mass by almost 18\%, and can produce very massive pulsars (2.2 $M_\odot$).
The chemical composition  also reflect the effect of strong magnetic fields.  This work is supported by CAPES, CNPq and FAPESC.

\end{document}